\documentclass[sigconf,natbib=true,anonymous=false]{acmart}

\usepackage{amssymb}
\usepackage{array}
\usepackage{url}
\usepackage{dblfloatfix}
\usepackage{booktabs}
\usepackage{amsmath}
\usepackage{tabularx}
\usepackage{colortbl}

\usepackage{graphics}
\usepackage{epsfig}
\usepackage{multirow}
\usepackage{flushend}

\usepackage{url}
\usepackage{tikz}
\usepackage{enumitem}
\usepackage{multicol}

\usepackage{verbatim}
\usepackage{float}
\usepackage{subcaption}

\usepackage{algorithm}
\usepackage{algpseudocode}

\usepackage{marginnote}
\usepackage{xspace}
\usepackage{balance}

\usepackage{todonotes}
\usepackage{xcolor}

\newcommand{\squishlist}{
 \begin{list}{$\bullet$}
  { \setlength{\itemsep}{0pt}
     \setlength{\parsep}{1pt}
     \setlength{\topsep}{1pt}
     \setlength{\partopsep}{0pt}
     \setlength{\leftmargin}{1.5em}
     \setlength{\labelwidth}{1em}
     \setlength{\labelsep}{0.5em} } }

\newcommand{\squishend}{
  \end{list}  }

\author{Iain Mackie}
\affiliation{
  \institution{University of Glasgow}
}
\email{i.mackie.1@research.gla.ac.uk}

\author{Shubham Chatterjee}
\affiliation{
  \institution{University of Glasgow}
}
\email{shubham.chatterjee@glasgow.ac.uk}

\author{Jeffrey Dalton}
\affiliation{
  \institution{University of Glasgow}
}
\email{jeff.dalton@glasgow.ac.uk}


\renewcommand\footnotetextcopyrightpermission[1]{} 

\begin{document}
\fancyhead{}

\title{Generative and Pseudo-Relevant Feedback for Sparse, \protect\\ Dense and Learned Sparse Retrieval}

\begin{abstract}

Pseudo-relevance feedback (PRF) is a classical approach to address lexical mismatch by enriching the query using first-pass retrieval. 
Moreover, recent work on generative-relevance feedback (GRF) shows that query expansion models using text generated from large language models can improve sparse retrieval without depending on first-pass retrieval effectiveness.
This work extends GRF to dense and learned sparse retrieval paradigms with experiments over six standard document ranking benchmarks.
We find that GRF improves over comparable PRF techniques by around 10\% on both precision and recall-oriented measures.
Nonetheless, query analysis shows that GRF and PRF have contrasting benefits, with GRF providing external context not present in first-pass retrieval, whereas PRF grounds the query to the information contained within the target corpus.
Thus, we propose combining generative and pseudo-relevance feedback ranking signals to achieve the benefits of both feedback classes, which significantly increases recall over PRF methods on 95\% of experiments.

\end{abstract}

\maketitle

%%%%%%%%%%%%%%%%%%%%%%%%%%%%%%%%%%%%%%%%%%%%%%%%%%%%%%%%%%%%%%%%%%%%%%%
%%%%%%%%%%%%%%%%%%%%%%%%%%%% NEW SECTION %%%%%%%%%%%%%%%%%%%%%%%%%%%%%%
%%%%%%%%%%%%%%%%%%%%%%%%%%%%%%%%%%%%%%%%%%%%%%%%%%%%%%%%%%%%%%%%%%%%%%%
\section{Introduction}
\label{sec:intro}

% -- PRF -- 
The traditional approach to address vocabulary mismatch~\cite{belkin1982ask} is  pseudo-relevance feedback (PRF)~\cite{abdul2004umass, zhai2001model, metzler2007latent, metzler2005markov}, where the query is expanded using information from the top-$k$ documents in a feedback set obtained using a first-pass retrieval. 
Several recent approaches leverage dense \cite{naseri2021ceqe, wang2022colbert, yu2021improving, li2022improving} and learned sparse representation~\cite{lassance2023naver} PRF to contextualise the query vector.
While PRF often improves recall, its effectiveness hinges on the quality of the first-pass retrieval. 
Non-relevant documents in the feedback set introduce noise and may push the query off-topic.  

% -- GRF -- 
Recently work by \citet{mackie2023generative} proposes generative-relevance feedback (GRF) that use Large Language Models (LLMs) to generate text independent of first-pass retrieval. 
Specifically, they use an LLM to generate diverse types of text based on the initial query and use these ``generated documents'' as input for term-based expansion models~\cite{abdul2004umass}.
They experiment using different types of generated content (chain-of-thought reasoning, facts, news articles, etc.) and find that combining multiple generations is best. 
We build upon this work to extend GRF to dense~\cite{li2022improving} and learned sparse~\cite{lassance2023naver} retrieval, where we encode generated text into dense and learned sparse vectors for query contextualisation.

\begin{figure}[h!]
    \centering
    \includegraphics[scale=0.19]{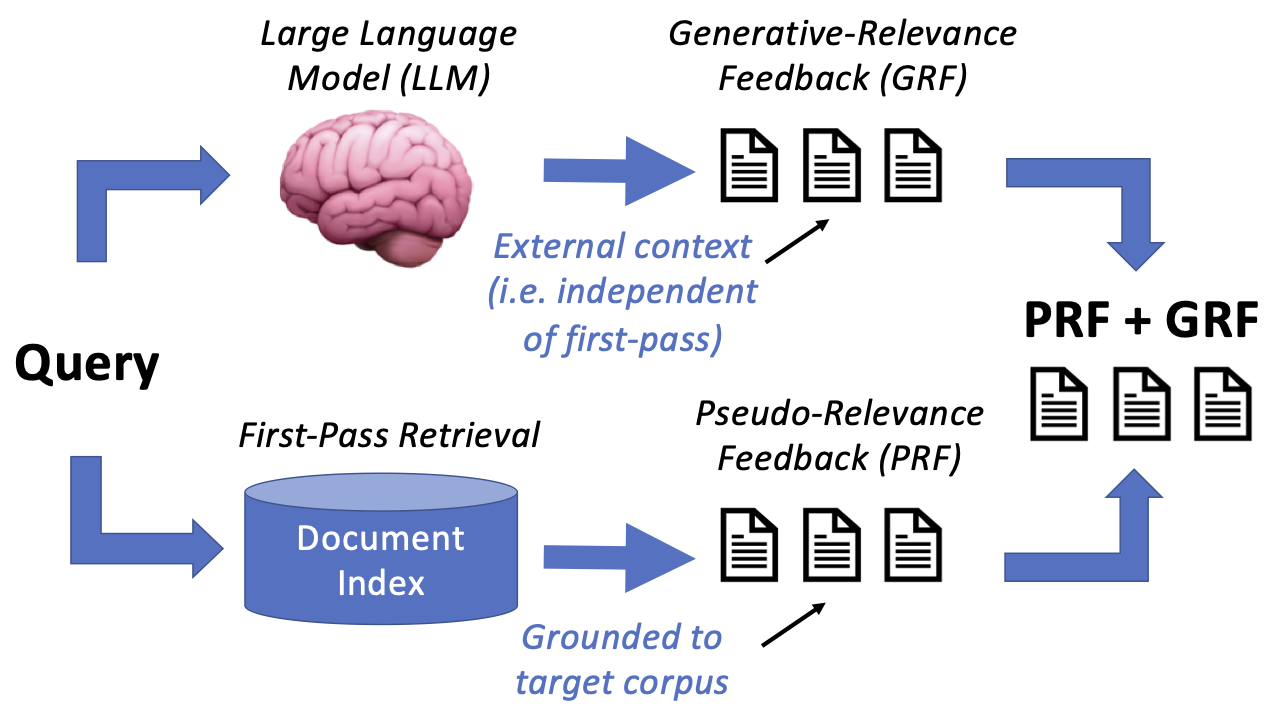}
    \caption{PRF+GRF: Combines generative and pseudo-relevance feedback to improve ranking effectiveness.}
    \label{img:grf+prf-overview}
\end{figure}

% PRF + GRF
Nonetheless, we conduct query analysis and find that generative and pseudo-relevance feedback have contrasting merits.
For example, GRF provides external context not present in first-pass retrieval, i.e. LLMs can explain how the practice of ``clear-cutting'' relates to [habitat], [climate] and [deforestation]. Conversely, LLMs can generate content not present in relevant documents. For example, PRF performs better on topics that need to be grounded to the corpus, i.e. ``human stampede'' where PRF correctly identifies events contained in relevant documents [Saudi] and [China], versus LLM-generated content that discusses stampedes in [India] and [Kerala] that are not present in relevant documents.
Based on this analysis, Figure~\ref{img:grf+prf-overview} shows our proposed weighted fusion method (PRF+GRF) that combines the ranking signals of PRF and GRF to improve recall-oriented effectiveness.
% Additionally, we propose default hyperparameters that are robust across datasets and search paradigms. 

% Experiments
For our experiments, we use six established document ranking benchmarks (see Section \ref{subsec:Datasets}) and the same GRF~\cite{mackie2023generative} LLM-generated content for consistency. 
We find dense and learned sparse retrieval with GRF improves over comparable PRF techniques on nDCG@20 by 9\% each, while recall improves for learned sparse by 8\% and dense by 11\%.
Furthermore, we show that combining GRF and PRF significantly improves R@1000. 

% Summary
Our results highlight how LLM-generated content can effectively contextualize queries across search paradigms (sparse, dense, learned sparse), and is consistently more effective than PRF. Furthermore, we show that generative and pseudo-relevant feedback ranking signals are complementary and can improve recall further.
To summarize the contributions, we provide the following:

\begin{itemize}[leftmargin=*]

\item We extend generative relevance feedback to dense and learned sparse retrieval, which improves precision and recall over comparable PRF models by around 10\%. 

\item We conduct qualitative and quantitative query analysis on PRF and GRF to show these techniques have contrasting benefits.

\item We show that GRF is complementary with PRF, and fusion significantly improves R@1000 across 17/18 experiments. 

\end{itemize}

\section{Related Work}
\label{sec:related-word}

\subsection{Pseudo-Relevance Feedback}
\label{sec:rw-prf}

% TERM PRF
The issue of vocabulary mismatch holds considerable importance in information retrieval, where a query fails to capture the user's entire information need~\cite{belkin1982ask}. Various techniques have been proposed to address this problem, including pseudo-relevance feedback (PRF)~\cite{rocchio1971relevance} that enriches the query with information from a first-pass set of candidate documents. This updated query is then re-issued to produce the second-pass retrieval set.
This process has been shown to improve recall, and famous examples include Rocchio~\cite{rocchio1971relevance}, KL expansion~\cite{zhai2001model}, relevance modelling~\cite{metzler2005markov}, LCE~\cite{metzler2007latent}, RM3~\cite{abdul2004umass}.
Furthermore, query expansion has been extended to include external corpora~\cite{diaz2006improving} and structured knowledge~\cite{meij2010conceptual, xiong2015query, Xu2009query, dalton2014entity}. 

% DENSE / LEARNED SPARSE PRF
The emergence of LLMs has shown progress across many different aspects of information retrieval~\cite{yates2021pretrained}.
One of the largest focus areas is dense retrieval~\cite{khattab2020colbert, lin2020distilling, xiongapproximate}, where passages are encoded into vectors and stored offline. Then, at query-time, the similarity is calculated between a query vector and the passage vectors. 
Vector-based pseudo-relevance feedback methods~\cite{li2022improving} use the first-pass vectors to transform the query vector. Examples include, ColBERT PRF~\cite{wang2022colbert}, ColBERT-TCT PRF~\cite{li2022improving}, and ANCE-PRF~\cite{yu2021improving}. Recently, learned sparse models like SPLADE~\cite{formal2021splade} use BERT~\cite{DBLP:conf/naacl/DevlinCLT19} and sparse regularization to learn query and document sparse weightings. 
Similarly, work has leveraged pseudo-relevance feedback with learned sparse representations~\cite{lassance2023naver} to improve retrieval effectiveness.
By contrast, our dense and learned sparse generative-relevance feedback approaches build feedback document vectors based on LLM-generation text embeddings. 

\subsection{Generative-Relevance Feedback}
\label{sec:rw-grf}

% \label{sec:rw-tg}
Large language models have been used to generate text to improve core retrieval effectiveness in several ways~\cite{jeronymo2023inpars, nogueira2019doc2query,  samarinas2022revisiting, macavaney2021intent5, hyde, liu2022query, pereira2023visconde, ferraretto2023exaranker, nogueira2019doc2query, wu-etal-2022-conqrr, zamani2020generating}.  
For instance, using LLMs for facet generation~\cite{macavaney2021intent5, samarinas2022revisiting} or document expansion via synthetic query generation~\cite{nogueira2019doc2query}. 
Recent work by \citet{bonifacio2022inpars} demonstrates the effectiveness of leveraging GPT3 for few-shot query generation to aid in dataset creation.
Furthermore, \citet{liu2022query} extracted contextual clues from LLMs, augmenting and merging multiple questions to improve QA effectiveness.
For passage ranking, HyDe~\cite{hyde} uses InstructGPT~\cite{ouyang2022training} to generate hypothetical document embeddings for dense retrieval.
Finally, notable works have focused on leveraging large language models for query-specific reasoning to improve ranking effectiveness~\cite{pereira2023visconde, ferraretto2023exaranker}.

Recent work proposes generative-relevance feedback (GRF)~\cite{mackie2023generative} that uses LLMs to generate query-specific text independent of first-pass retrieval. 
Specifically, GPT-3~\cite{brown2020language} is prompted to generate ten diverse types of text that act as ``generated documents'' for term-based expansion models~\cite{abdul2004umass}.
They experiment using different types of generated content (chain-of-thought reasoning, facts, news articles, etc.) and find that aggregating multiple generated documents is most effective. 
We build upon this work to extend GRF to dense~\cite{li2022improving} and learned sparse~\cite{lassance2023naver} retrieval paradigms, where we encode generated content into dense and learned sparse vectors for query contextualisation.
Furthermore, we conduct qualitative and quantitative analyses on the retrieval benefits of GRF versus PRF, proposing a fusion method to combine these complementary ranking signals.

\section{Method}
\label{sec:method}

In this section, we extend generative-relevance feedback to dense~\cite{Li2021PseudoRF} and learned sparse~\cite{formal2021splade} retrieval paradigms. 
GRF tackles query-document lexical mismatch using zero-shot LLM text generation.
Unlike traditional PRF approaches for query expansion \cite{abdul2004umass, metzler2005markov, metzler2007latent}, GRF is not reliant on first-pass retrieval effectiveness to find feedback documents.
Instead, we leverage LLMs \cite{brown2020language} to generate zero-shot relevant text content.

Furthermore, based on our analysis in Section~\ref{sec:rq2}, we propose combining GRF and PRF ranking signals to further improve retrieval effectiveness. 
Specifically, we find that GRF is useful provides external context and not dependent on first-pass retrieval effectiveness, while PRF grounds the query in corpus-specific information.
Therefore, we propose a weighted reciprocal rank fusion method that achieves the benefits of both feedback classes.
% Based on the findings in Section~\ref{sec:rq3}, we propose default relative weightings for GRF and PRF that achieves strong recall-oriented effectiveness across datasets and models.

\subsection{Extending GRF to Dense and Learned Sparse}

We extend generative relevance feedback to dense and learned sparse retrieval, focusing on the full ``GRF'' variant that combines multiple generated text contents.
Formally, for a given query $Q$, we want to return a ranked list of document $[D_1, D_2, ..., D_N]$ from a corpus $C$.
We have a Large Language Model, $LLM$, that generates $k$ number of documents $[D_{LLM_1}, D_{LLM_2}, ..., D_{LLM_k}]$.
Unlike sparse GRF~\cite{mackie2023generative}, where text can be aggregated into a single ``generated document'' to create a language model, dense~\cite{lin2021batch} and learned sparse~\cite{formal2021splade} embedding models are limited to a maximum number of input tokens. Thus, we treat each generated document as a separate dense or learned sparse embedding and combined post-embedding.     

\subsubsection{Sparse GRF}
\label{sec:sparse-grf}

See the work by \cite{mackie2023generative}. We follow similar notation for dense and sparse formulations.

\subsubsection{Dense GRF}
\label{sec:dense-grf}

We adopt the Rocchio PRF approach~\cite{li2022improving} for dense GRF to allow different weighting of the query vector and the feedback vector. This allows embeddings of LLM-generated text to contextualise the query vector in a controllable way. Specifically, Equation~\ref{eq:roccio-dense} shows that the new vector, $\vec{GRF}$, is the combination of the original query vector, $\vec{Q}$, and mean of the generated document vectors, $\vec{D_{LLM}} = 1/k \times (\vec{D_{LLM_1}} + \vec{D_{LLM_2}} + ... + \vec{D_{LLM_k}})$. We include $\alpha$ and $\beta$ to weigh the relative importance of query and GRF vectors. 

\begin{equation}
  \centering
   \label{eq:roccio-dense}
  \vec{GRF} = \alpha \vec{Q} + 
    \beta \vec{D_{LLM}}
\end{equation}

\subsubsection{Learned Sparse GRF}
\label{sec:LS-grf}

We draw on prior work combining pseudo-relevance feedback with learned sparse representations~\cite{lassance2023naver}.
Specifically, we combine our normalised learned sparse representations of the query, $LS(w|Q)$, with the representation of the combined generated documents, $LS(w|D_{LLM})$.
For $LS(w|D_{LLM})$, similar to \cite{nguyen2023adapting}, we normalise each generated document's sparse representation and aggregate them together before normalising again. 
$\beta$ (original query weight) is a hyperparameter to weigh the relative importance of our generative learned expansion terms, and $\theta$ (number of expansion terms) limits the most probable LLM-generated learned sparse representations.
This results in the weightings of the query learned sparse to be re-weighted and new terms added based on the generated documents.  

\begin{equation}
  \centering
   \label{eq:prob-grf}
  LS_{GRF}(w|R) = \beta LS(w|Q) + \begin{cases}
    (1 - \beta) LS(w|D_{LLM}), & \text{if $w \in W_{\theta}$}.\\
    0, & \text{otherwise}.
  \end{cases}
\end{equation}

\subsection{Fusion of GRF and PRF}
\label{sec:prf-grf-fusion}

Query analysis in Section~\ref{sec:rq2} shows that generative and pseudo-relevance feedback have different retrieval benefits. 
Specifically, generative feedback can provide external context without being dependent on first-pass effectiveness, while  pseudo-relevance feedback can provide corpus-specific contextualization. 
Thus, we propose combing these document scoring signals.
Specifically, Equation \ref{eq:wrrf} shows our weighted reciprocal rank fusion method (WRRF) (adapted from ~\cite{cormack2009reciprocal}) that combines our GRF and PRF runs (PRF+GRF).
Here, WRRF uses a scoring formula, $r(d)$, based on the document's rank in a specific run. There is a set $D$ of documents to be ranked, a set of rankings $R$, and $k$ parameter is included so low-rank document signals do not disappear (default usual 60).
We add a hyperparameter $\lambda$, which weights the relative importance of pseudo-relevant document rankings, $r \in R_{PRF}$, and $(1-\lambda)$ for generative document rankings, $r \in R_{PRF}$.
This formulation allows us to tune the relative weighting of GRF and PRF across models and datasets in Section \ref{sec:rq3}.

\begin{equation}
  \centering
   \label{eq:wrrf}
    WRRF(d \in D) = \sum_{r \in R} 1 / (k + r(d)) \times \begin{cases}
    \lambda, & \text{if $r \in R_{PRF}$}.\\
    (1 - \lambda), & \text{if $r \in R_{GRF}$}.
  \end{cases} 
\end{equation}

\section{Experimental Setup}
\label{sec:exp-setup}

\subsection{Datasets}
\label{subsec:Datasets}

\subsubsection{Retrieval Corpora}

We evaluate using three test collections and six query sets, which provide a diverse evaluation: 
%This includes different types of corpora, i.e. newswire, domain-focused and web document, and different types of queries, i.e. challenging, essay-style and factoid-based queries.     

\textbf{TREC Robust04} \cite{Voorhees_TREC2004_robust}: TREC 2004 Robust Track was created to target poorly performing topics. 
This dataset comprises 249 topics, containing short keyword ``titles'' and longer natural-language "descriptions" queries. 
Scaled relevance judgments are over a newswire collection of 528k long documents (TREC Disks 4 and 5).

\textbf{TREC Deep Learning 19/20/HARD}: The MS MARCO document collections~\cite{nguyen2016ms} consist of queries, web documents, and sparse relevance judgments. 
The TREC Deep Learning (DL) 19/20 document track~\cite{craswell2020overview, craswell2021overview} builds upon the MS MARCO collection and NIST provides judgments pooled to a greater depth, containing 43 topics for DL-19 and 45 topics for DL-20. 
DL-HARD~\cite{mackie2021deep} provides a challenging subset of 50 queries focusing on topics requiring reasoning and having multiple answers.

\textbf{CODEC}~\cite{mackie2022codec} is a dataset that focuses on the complex information needs of social science researchers, where domain experts (economists, historians, and politicians) generate 42 challenging essay-style topics.
CODEC has a focused web corpus of 750k long documents, which includes news (BBC, Reuters, CNBC etc.) and essay-based web content (Brookings, Forbes, eHistory, etc.).

\subsubsection{Indexing and Evaluation} For indexing the corpora we use Pyserini version 0.16.0 \cite{lin2021pyserini}, removing stopwords and using Porter stemming. 
We use cross-validation and optimise R@1000 on standard folds for Robust04~\cite{huston2014parameters}, CODEC~\cite{mackie2022codec}, and DL-HARD~\cite{mackie2021deep}. On DL-19, we cross-validated on DL-20 and use the average parameters zero-shot on DL-19 (and vice versa for DL-20). 

We assess the system runs to a run depth of 1,000. 
Furthermore, because GRF is an initial retrieval model, recall-oriented evaluation is important; Recall@1000 is the primary measure for this paper. Nonetheless, we include MAP and nDCG@10 for an understanding of precision. 
We evaluate using ir-measures~\cite{macavaney2022streamlining} and conduct 95\% confidence paired-t-test for significance.

\begin{table*}[h!] 
% \small
\caption{Dense and learned sparse generative relevance feedback. ``+'' indicates significant improvements in dense systems against 
TCT+PRF and learned sparse systems against SPLADE+RM3, with \textit{bold} depicting the best system.}
\tabcolsep=0.12cm
\label{tab:content-dense-LS}
\begin{tabular}{lllllllllllll}
\cline{2-13}
\multicolumn{1}{l|}{}                        & \multicolumn{3}{c|}{Robust04 -Title}                                                            & \multicolumn{3}{c|}{Robust04 - Descriptions}                                                    & \multicolumn{3}{c|}{DL-19}                                                                      & \multicolumn{3}{c|}{DL-20}                                                                      \\ \cline{2-13} 
\multicolumn{1}{l|}{\textit{\textbf{Dense}}} & \multicolumn{1}{c}{nDCG@10} & \multicolumn{1}{c}{MAP} & \multicolumn{1}{c|}{R@1k}               & \multicolumn{1}{c}{nDCG@10} & \multicolumn{1}{c}{MAP} & \multicolumn{1}{c|}{R@1k}               & \multicolumn{1}{c}{nDCG@10} & \multicolumn{1}{c}{MAP} & \multicolumn{1}{c|}{R@1k}               & \multicolumn{1}{c}{nDCG@10} & \multicolumn{1}{c}{MAP} & \multicolumn{1}{c|}{R@1k}               \\ \hline
\multicolumn{1}{|l|}{ColBERT}                & 0.445                       & 0.233                   & \multicolumn{1}{l|}{0.608}              & 0.435                       & 0.218                   & \multicolumn{1}{l|}{0.605}              & 0.634                       & 0.320                   & \multicolumn{1}{l|}{0.564}              & 0.611                       & 0.429                   & \multicolumn{1}{l|}{0.795}              \\
\multicolumn{1}{|l|}{ColBERT+PRF}            & 0.467                       & 0.272                   & \multicolumn{1}{l|}{0.648}              & 0.461                       & 0.263                   & \multicolumn{1}{l|}{0.635}              & 0.668                       & 0.385                   & \multicolumn{1}{l|}{0.625}              & 0.615                       & \textbf{0.489}          & \multicolumn{1}{l|}{\textbf{0.813}}     \\
\multicolumn{1}{|l|}{TCT}                    & 0.466                       & 0.233                   & \multicolumn{1}{l|}{0.637}              & 0.424                       & 0.214                   & \multicolumn{1}{l|}{0.595}              & 0.655                       & 0.333                   & \multicolumn{1}{l|}{0.638}              & 0.600                       & 0.412                   & \multicolumn{1}{l|}{0.771}              \\
\multicolumn{1}{|l|}{TCT+PRF}                & 0.493                       & 0.274                   & \multicolumn{1}{l|}{0.684}              & 0.452                       & 0.245                   & \multicolumn{1}{l|}{0.628}              & 0.670                       & 0.378                   & \multicolumn{1}{l|}{0.684}              & 0.618                       & 0.442                   & \multicolumn{1}{l|}{0.784}              \\
\multicolumn{1}{|l|}{TCT+GRF}                    & \textbf{0.517$^+$}          & \textbf{0.276}          & \multicolumn{1}{l|}{\textbf{0.700$^+$}} & \textbf{0.571$^+$}          & \textbf{0.289$^+$}      & \multicolumn{1}{l|}{\textbf{0.708$^+$}} & \textbf{0.683}              & \textbf{0.418$^+$}      & \multicolumn{1}{l|}{\textbf{0.743$^+$}} & \textbf{0.634}              & 0.457                   & \multicolumn{1}{l|}{0.812$^+$}          \\ \hline
\textit{\textbf{Learned Sparse}}             &                             &                         &                                         &                             &                         &                                         &                             &                         &                                         &                             &                         &                                         \\ \hline
\multicolumn{1}{|l|}{SPLADE}                 & 0.387                       & 0.206                   & \multicolumn{1}{l|}{0.660}              & 0.426                       & 0.230                   & \multicolumn{1}{l|}{0.672}              & 0.552                       & 0.280                   & \multicolumn{1}{l|}{0.619}              & \textbf{0.553}                       & 0.352                   & \multicolumn{1}{l|}{0.779}              \\
\multicolumn{1}{|l|}{SPLADE+RM3}             & 0.418                       & 0.248                   & \multicolumn{1}{l|}{0.703}              & 0.448                       & 0.268                   & \multicolumn{1}{l|}{0.715}              & 0.566                       & 0.328                   & \multicolumn{1}{l|}{0.651}              & 0.533                       & 0.379                   & \multicolumn{1}{l|}{0.784}              \\
\multicolumn{1}{|l|}{SPLADE+GRF}                    & \textbf{0.462$^+$}          & \textbf{0.265$^+$}      & \multicolumn{1}{l|}{\textbf{0.730$^+$}} & \textbf{0.493$^+$}          & \textbf{0.276}          & \multicolumn{1}{l|}{\textbf{0.732$^+$}} & \textbf{0.642$^+$}          & \textbf{0.407$^+$}      & \multicolumn{1}{l|}{\textbf{0.732$^+$}} & \textbf{0.553}              & \textbf{0.415}          & \multicolumn{1}{l|}{\textbf{0.839$^+$}} \\ \hline
\end{tabular}
\end{table*}

\subsection{GRF Implementation}

\subsubsection{LLM Generation}

For a fairness and reproducible, we use the same generated content as \cite{mackie2023generative}: \href{https://drive.google.com/drive/folders/1LWGTvXGatrAbwbDahYkraK-nim2O9eyN?usp=sharing}{\textbf{{\textit{link}}}}.

\subsubsection{Retrieval and GRF} We outline the different sparse, dense and learned sparse implementation details for GRF:

\noindent \textbf{BM25+GRF}: We use the runs provided by \cite{mackie2023generative}.

\noindent \textbf{TCT+GRF}:
We use the TCT-ColBERT-v2-HNP's~\cite{Li2021PseudoRF} model trained on MS MARCO \cite{MS_MARCO_v1}. We shard documents into passages of 10 sentences, encoding the document title within each passage, and use a stride length of 5. A max-passage approach transforms passage scores into document scores.
We use the ColBERT-TCT encoder to create the GRF document vectors, and we tune the Rocchio PRF $\alpha$ (between 0.1 and 0.9 with a step of 0.1, and $\beta$ (between 0.1 and 0.9 with a step of 0.1). 

\noindent \textbf{SPLADE+GRF}: 
We use the SPLADE \cite{formal2021splade} \textit{naver/splade-cocondenser-ensembledistil} checkpoint to create a passage index using the same processing as TCT+GRF. We index the term vectors using Pyserini~\cite{lin2021pyserini} and use their ``impact'' searcher for max-passage aggregation. When combining query or document vectors we normalise the weights of each term. We tune $fb\_terms$ (20,40,60,80,100) and $original\_query\_weight$ (between 0.1 and 0.9 with a step of 0.1).

\subsection{Comparison Methods}

\noindent \textbf{BM25} \cite{robertson1994some}: 
Sparse retrieval method where $k1$ parameter was tuned between 0.1 and 5.0 using a step size of 0.2, while $b$ was tuned between 0.1 and 1.0 with a step size of 0.1, as described earlier.

\noindent \textbf{BM25+Relevance Model (RM3)} \cite{abdul2004umass}: 
We tune $fb\_terms$ (between 10 and 100 with a step of 10), $fb\_docs$ (between 10 and 100 with a step of 10), and $original\_query\_weight$ (between 0.1 and 0.9 with a step of 0.1). 
%This is the primary baseline for GRF.

% \noindent  \textbf{CEQE} \cite{naseri2021ceqe}: Utilizes query-focused contextualized embedding vectors for query expansion. For Robust04 we use the CEQE-MaxPool runs provided by the author, and for DL-19 and Dl-20 we use the CEQE runs provided by \cite{wang2023colbert}. 

\noindent  \textbf{ColBERT-TCT (TCT)} \cite{lin2021batch}: is a dense retrieval model incorporating knowledge distillation over ColBERT \cite{khattab2020colbert}. We employ TCT-ColBERT-v2-HNP's MS MARCO \cite{MS_MARCO_v1} model and use a max-passage approach to convert our passage runs into document runs. For \textbf{ColBERT-TCT+PRF (TCT+PRF)}~\cite{Li2021PseudoRF},
we tune Rocchio PRF parameters: $depth$ (2,3,5,7,10,17), $\alpha$ (between 0.1 and 0.9 with a step of 0.1, and $\beta$ (between 0.1 and 0.9 with a step of 0.1). 

\noindent  \textbf{SPLADE} \cite{formal2021splade}: is a neural retrieval model which learns sparse query and document weightings via the BERT MLM head and sparse regularization. We index the term vectors using Pyserini \cite{lin2021pyserini} and use their ``impact'' searcher for max-passage aggregation. For \textbf{SPLADE+RM3}, we tune $fb\_docs$ (5,10,15,20,25,30) $fb\_terms$ (20,40,60,80,100), and $original\_query\_weight$ (between 0.1 and 0.9 with a step of 0.1).

\noindent \textbf{ColBERT \cite{khattab2020colbert} \& ColBERT+PRF}~\cite{wang2022colbert}: We use the runs provided by~\citet{wang2023colbert}, which use pyterrier framework~\cite{macdonald2021pyterrier}.

\subsection{Fusing GRF and PRF runs}

We use weighted reciprocal rank fusion (WRRF) as a simple fusion method. 
We extend TrecTools~\cite{palotti2019}  RRF~\cite{cormack2009reciprocal} implementation and keep $k=60$ as default. 
We tune $\lambda$ between 0.0 and 1.0 with a step size of 0.1. This hyperparameter does change across datasets and models, although typically within the 0.2-0.5 range, i.e., slightly favouring GRF. Furthermore, we notice $\lambda$ is relatively consistent across folds (i.e. +/- 0.1 at most).
We fuse GRF and PRF runs within the same search paradigm:
\begin{itemize}[leftmargin=*]

\item \textbf{BM25+PRF+GRF}: BM25+RM3 with BM25+GRF.

\item \textbf{TCT+PRF+GRF}: TCT+PRF with TCT+GRF.

\item \textbf{SPLADE+PRF+GRF}: SPLADE+RM3 with SPLADE+GRF.

\end{itemize}

\section{Results \& Analysis}
\label{sec:results}

\subsection{Research Questions}

\begin{itemize}[leftmargin=*]

\item \textbf{RQ1}: \textit{Does generative-relevance feedback improve the effectiveness of dense and learned sparse models?} We extend GRF to new search paradigms and show the effectiveness gains.

\item \textbf{RQ2}: \textit{What queries does generative-relevance feedback impact versus traditional pseudo-relevance feedback?} We conduct query analysis to understand the behaviours of GRF and PRF methods.

\item \textbf{RQ3}: \textit{Do generative and pseudo-relevance feedback methods have complementary ranking signals?} We explore combing GRF and PRF using our fusion method.

\end{itemize}

\subsection{RQ1: Dense and Learned Sparse GRF}
\label{sec:rq1}

% --- table overview ---
Table~\ref{tab:content-dense-LS} shows the effectiveness of generative feedback with dense and learned sparse retrieval on Robust04 and DL datasets.
We test for significant improvements against ColBERT-TCT PRF for our dense table section and SPLADE with RM3 expansion for the learned sparse table section.

% --- Dense + LS---
ColBERT-TCT with GRF improves, often significantly, across all datasets. For example, we improve R@1000 over PRF between 2-13\% and show significant improvements on all datasets. 
Additionally, GRF improves nDCG@10 between 2-27\% and significantly on 3/4 datasets, i.e. DL-19 and Robust04 titles and descriptions. 
Furthermore, ColBERT-TCT with GRF outperforms the full late interaction ColBERT-PRF model on all datasets and measures, except underperforming recall-oriented measures on DL-20.  
Similarly, SPLADE with GRF improves R@1000 over PRF between 2-12\% and shows significant improvements on all datasets. 
Additionally, we see improvements in nDCG@10 between 4-13\% and significantly on 3/4 datasets, i.e. DL-19 and Robust04 titles and descriptions. 

% --- Overview RQ ---
These results, combined with sparse GRF findings~\cite{mackie2023generative}, support GRF as an effective and robust query augmentation approach across sparse, dense and learned sparse retrieval paradigms.
In the next research question, we conduct query analysis to understand the different behaviour of generated and pseudo-relevant feedback.  

\begin{table*}[b!]
\caption{The effectiveness of fusing PRF and GRF runs within each search paradigm. ``+'' indicates significant improvements against PRF from the respective search paradigm (i.e., BM25+RM3 for sparse, etc.), and \textit{bold} depicts best system.}
\centering
\label{tab:fuse-search}
\begin{tabular}{lllllllllllll}
\cline{2-13}
\multicolumn{1}{l|}{}                         & \multicolumn{2}{c|}{Robust04 - Title}                        & \multicolumn{2}{c|}{Robust04 - Desc}                         & \multicolumn{2}{c|}{CODEC}                                          & \multicolumn{2}{c|}{DL-19}                                   & \multicolumn{2}{c|}{DL-20}                                   & \multicolumn{2}{c|}{DL-HARD}                                        \\ \cline{2-13} 
\multicolumn{1}{c|}{\textit{\textbf{Sparse}}} & MAP                & \multicolumn{1}{l|}{R@1k}               & MAP                & \multicolumn{1}{l|}{R@1k}               & MAP                       & \multicolumn{1}{l|}{R@1k}               & MAP                & \multicolumn{1}{l|}{R@1k}               & MAP                & \multicolumn{1}{l|}{R@1k}               & MAP                       & \multicolumn{1}{l|}{R@1k}               \\ \hline
\multicolumn{1}{|l|}{BM25+RM3}                & 0.292              & \multicolumn{1}{l|}{0.777}              & 0.278              & \multicolumn{1}{l|}{0.750}              & \multicolumn{1}{c}{0.239} & \multicolumn{1}{c|}{0.816}              & 0.383              & \multicolumn{1}{l|}{0.745}              & 0.418              & \multicolumn{1}{l|}{0.825}              & \multicolumn{1}{c}{0.190} & \multicolumn{1}{c|}{0.787}              \\
\multicolumn{1}{|l|}{BM25+GRF}                & 0.307              & \multicolumn{1}{l|}{0.788}              & 0.318$^+$          & \multicolumn{1}{l|}{0.776$^+$}          & \textbf{0.285$^+$}        & \multicolumn{1}{l|}{0.830}              & 0.441$^+$          & \multicolumn{1}{l|}{0.797$^+$}          & \textbf{0.486$^+$} & \multicolumn{1}{l|}{0.879$^+$}          & 0.241$^+$                 & \multicolumn{1}{l|}{0.817}              \\
\multicolumn{1}{|l|}{BM25+PRF+GRF}            & \textbf{0.323$^+$} & \multicolumn{1}{l|}{\textbf{0.817$^+$}} & \textbf{0.331$^+$} & \multicolumn{1}{l|}{\textbf{0.823$^+$}} & 0.275$^+$                 & \multicolumn{1}{l|}{\textbf{0.853$^+$}} & \textbf{0.442$^+$} & \multicolumn{1}{l|}{\textbf{0.803$^+$}} & 0.484$^+$          & \multicolumn{1}{l|}{\textbf{0.889$^+$}} & \textbf{0.243$^+$}        & \multicolumn{1}{l|}{\textbf{0.828$^+$}} \\ \hline
\textit{\textbf{Dense}}                       &                    &                                         &                    &                                         &                           &                                         &                    &                                         &                    &                                         &                           &                                         \\ \hline
\multicolumn{1}{|l|}{TCT+PRF}                 & 0.274              & \multicolumn{1}{l|}{0.684}              & 0.245              & \multicolumn{1}{l|}{0.628}              & 0.239                     & \multicolumn{1}{l|}{0.757}              & 0.378              & \multicolumn{1}{l|}{0.684}              & 0.442              & \multicolumn{1}{l|}{0.784}              & 0.228                     & \multicolumn{1}{l|}{0.745}              \\
\multicolumn{1}{|l|}{TCT+GRF}                 & 0.276              & \multicolumn{1}{l|}{0.700$^+$}          & 0.289$^+$          & \multicolumn{1}{l|}{0.708$^+$}          & \textbf{0.261}            & \multicolumn{1}{l|}{\textbf{0.821$^+$}} & \textbf{0.418$^+$} & \multicolumn{1}{l|}{\textbf{0.743$^+$}} & 0.457              & \multicolumn{1}{l|}{0.812$^+$}          & 0.228                     & \multicolumn{1}{l|}{0.757}              \\
\multicolumn{1}{|l|}{TCT+PRF+GRF}             & \textbf{0.287$^+$} & \multicolumn{1}{l|}{\textbf{0.707$^+$}} & \textbf{0.303$^+$} & \multicolumn{1}{l|}{\textbf{0.727$^+$}} & \textbf{0.261}            & \multicolumn{1}{l|}{\textbf{0.821$^+$}} & 0.416$^+$          & \multicolumn{1}{l|}{0.742$^+$}          & \textbf{0.464$^+$} & \multicolumn{1}{l|}{\textbf{0.814$^+$}} & \textbf{0.234}            & \multicolumn{1}{l|}{\textbf{0.764$^+$}} \\ \hline
\textit{\textbf{Learned Sparse}}              &                    &                                         &                    &                                         &                           &                                         &                    &                                         &                    &                                         &                           &                                         \\ \hline
\multicolumn{1}{|l|}{SPLADE+RM3}              & 0.248              & \multicolumn{1}{l|}{0.703}              & 0.268              & \multicolumn{1}{l|}{0.715}              & 0.216                     & \multicolumn{1}{l|}{0.770}              & 0.328              & \multicolumn{1}{l|}{0.651}              & 0.379              & \multicolumn{1}{l|}{0.784}              & 0.157                     & \multicolumn{1}{l|}{0.704}              \\
\multicolumn{1}{|l|}{SPLADE+GRF}              & 0.265$^+$          & \multicolumn{1}{l|}{0.730$^+$}          & \textbf{0.276}     & \multicolumn{1}{l|}{0.732$^+$}          & 0.222                     & \multicolumn{1}{l|}{0.785}              & \textbf{0.407$^+$} & \multicolumn{1}{l|}{\textbf{0.732$^+$}} & 0.415              & \multicolumn{1}{l|}{0.839$^+$}          & \textbf{0.182}            & \multicolumn{1}{l|}{\textbf{0.758$^+$}} \\
\multicolumn{1}{|l|}{SPLADE+PRF+GRF}          & \textbf{0.265$^+$} & \multicolumn{1}{l|}{\textbf{0.743$^+$}} & \textbf{0.276$^+$} & \multicolumn{1}{l|}{\textbf{0.757$^+$}} & \textbf{0.225}            & \multicolumn{1}{l|}{\textbf{0.790}}     & 0.401$^+$          & \multicolumn{1}{l|}{\textbf{0.732$^+$}} & \textbf{0.420$^+$} & \multicolumn{1}{l|}{\textbf{0.840$^+$}} & \textbf{0.182}            & \multicolumn{1}{l|}{\textbf{0.758$^+$}} \\ \hline
\end{tabular}
\end{table*}

%%%%%%%%%%%%%%%%%%%%%%%%%%%%%%%%%%%%%%%%%%%%%%%%%%%%%%%%%%%
\subsection{RQ2: Query Analysis of Different Feedback}
\label{sec:rq2}

\begin{figure}[h!]
    \centering
    \includegraphics[scale=0.43]{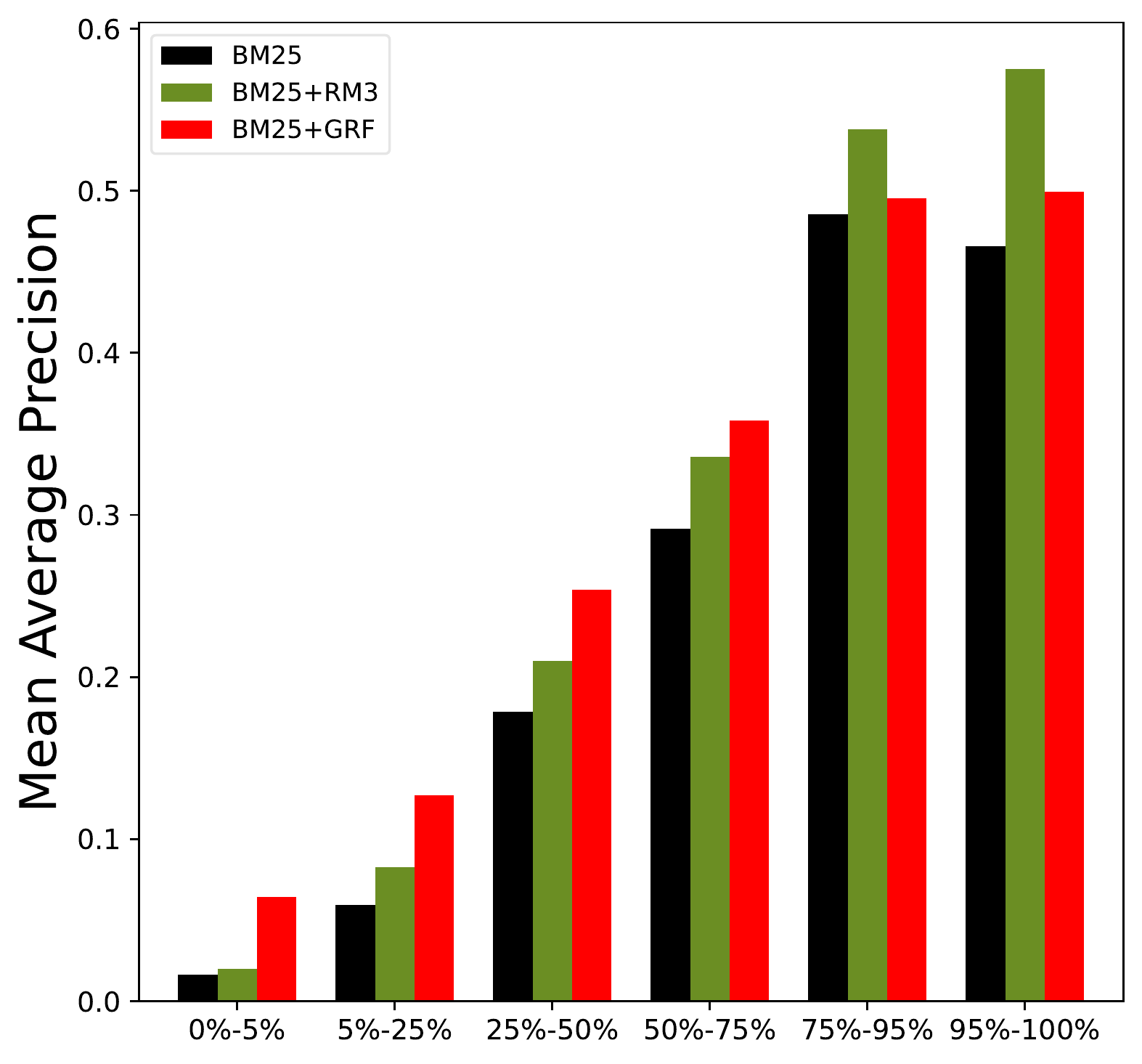}
    \caption{Query difficulty plot stratified by nDCG@10 of BM25 on Robust04 titles. We show MAP effectiveness of BM25 and BM25 with RM3 and GRF expansion.
    }
    \label{img:diff-plot}
\end{figure}

% -- chart --
In this section, we analyse the sparse GRF runs~\cite{mackie2023generative}, as well are our dense and learned sparse extensions.
Figure~\ref{img:diff-plot} shows the query difficulty plot stratified by nDCG@10 effectiveness of BM25 on Robust04 titles.
We report MAP and also include BM25 with RM3 and GRF expansion.
Specifically, this shows the hardest (0-25\%) and easiest (75-100\%) first-pass queries based on precision.
It is noticeable that GRF is better than PRF on the hardest 75\% of first-pass queries (0-75\% in our chart), with MAP consistently above RM3 expansion. 
However, on the easiest first-pass queries (75-100\% in our chart), RM3 is more effective than GRF.
In essence, we show that pseudo-relevance feedback is more effective than generative-relevance feedback when first-pass precision is very high. 

% --- Hurts vs helps ---
Furthermore, we analyse the query helps vs hurts for Robust04 titles, comparing BM25 query effectiveness to RM3 and GRF expansion.
For R@1000, RM3 hurts 47 queries and helps 139 queries, while GRF impacts more queries, hurting 53 and helping 150.
Interestingly, of the 47 queries that RM3 hurts, 33 (70\%) are either improved or unaffected by GRF. 
Conversely, of the 53 queries where GRF hurts effectiveness, 40 (75\%) of those queries are helped or unaffected by RM3 expansion.
Again, this highlights that generative and pseudo-relevance feedback affect different queries, suggesting these could have complementary ranking signals.

% --- Hard first pass ---
Specifically, a hard first-pass topic is 691 on Robust04 descriptions, \textit{What are the objections to the practice of ``clear-cutting''}. BM25 has a R@1000 of 0.333 and nDCG@10 of 0.000, and RM3 expansion further reduces R@1000 to 0.286 with nDCG@10 unchanged at 0.000. 
Reviewing the pseudo-relevant documents used for RM3 expansion, these are general documents around the forest industry and expand with terms: [forest], [timber], [logging], and [industry].
On the other hand, GRF expansion uses generated content that directly addresses the questions and injects terms about the ``objections'' to clear-cutting. Analysing the different generated documents \cite{mackie2023generative}, CoT-Keywords expands with [habitat], [climate] and [deforestation], Facts expands with [flooding], [water], [risk], and News expands with [soil], [erosion], and [environment].
This results in BM25 with GRF expansion increasing R@1000 to 0.810 and nDCG@10 to 0.454. 

% --- Easy first pass ---
In contrast, an easy query for first-pass retrieval is topic 626 on Robust04 titles, \textit{human stampede}, where BM25 achieves nDCG@10 of 0.287 and R@1000 of 1.000, and RM3 expansion improves nDCG@10 to 0.517 while retaining perfect recall. Reviewing the RM3 expansion terms and the relevant documents, we see specific terms that are useful and refer to collection-mention stampedes, i.e., [Saudi], [China], [pilgrimage], and [Calgary]. Conversely, without being grounded in the events covered in the collection, GRF expands with general terms, i.e. [human], [death], [crowd], [panic], [tragedy], or terms relating to events not converted in the collection, i.e. [India], [Kerala], [2011], etc. In this case, the LLM-generated content refers to the Sabarimala Temple stampede, which occurred seven years after the Robust04 corpus. 

% --- Dense entity centric-queries ---
Lastly, Dense retrieval has been shown to fail on even simple entity-centric queries~\cite{sciavolino2021simple}. Nonetheless, we observe that GRF can inject external context to help improve the effectiveness of this query type. For example, on topic 1103812 of DL-19, \textit{who formed the Commonwealth of Independent States},  ColBERT-TCT has a R@1000 of 0.425 and nDCG@10 of 0.574, while PRF slightly improves recall to 0.6000 but reduced nDCG@10 to 0.568.
However, we find that LLM-generated content explicitly provides entities to answer the question, i.e. discusses the Belavezha Accords and the three founding states, i.e. [Russia], [Ukraine], and [Belarus], and the Alma-Ata Protocol, which extends the Commonwealth to include [Armenia], [Azerbaijan], [Kazakhstan], [Kyrgyzstan], [Moldova], [Tajikistan], [Turkmenistan], and [Uzbekistan]. These different GRF embedding results in ColBERT-TCT with GRF improving R@1000 of 0.8000 and nDCG@10 of 0.705.

Overall, this analysis highlights that generative and pseudo-relevance feedback help different profiles of queries, which could suggest they are complementary.
Thus, in the next research question, we explore combing the ranking signals of generative and pseudo-relevant feedback.

%%%%%%%%%%%%%%%%%%%%%%%%%%%%%%%%%%%%%%%%%%%%%%%%%%%%%%%%%%%
\subsection{RQ3: Combining GRF and PRF}
\label{sec:rq3}

% --- overview of table ---
Table~\ref{tab:fuse-search} shows the effectiveness of PRF, GRF and our weighted reciprocal rank fusion (PRF+GRF) across our three search paradigms (sparse, dense, learned sparse).
For sparse, we use BM25 with RM3 expansion as the baseline for significance testing; for dense, we use ColBERT-TCT with PRF; for learned sparse, we use SPLADE with RM3 expansion.

% --- overview of R@1000 ---
We find that combining PRF and GRF consistently, often significantly, improves recall across datasets and search paradigms. 
For example, fusion has the best R@1000 across 14/18 and significantly improves over PRF on 17/18 experiments (4 more than GRF alone). 
We find consistent improvements in R@1000 across the search paradigms over PRF, with fusion increasing sparse by 6.7\%, dense by 8.7\%, and learned sparse by 6.9\%.
Furthermore, fusion shows consistent improvement over GRF, increasing sparse by 2.6\%, dense by 2.3\%, and learned sparse by 1.0\%.
Nonetheless, because GRF has much higher precision effectiveness when compared to PRF, nDCG@10 can be negatively impacted, and we do not see the same gains as recall-oriented evaluation.

\begin{figure}[h!]
    \centering
    \includegraphics[scale=0.4]{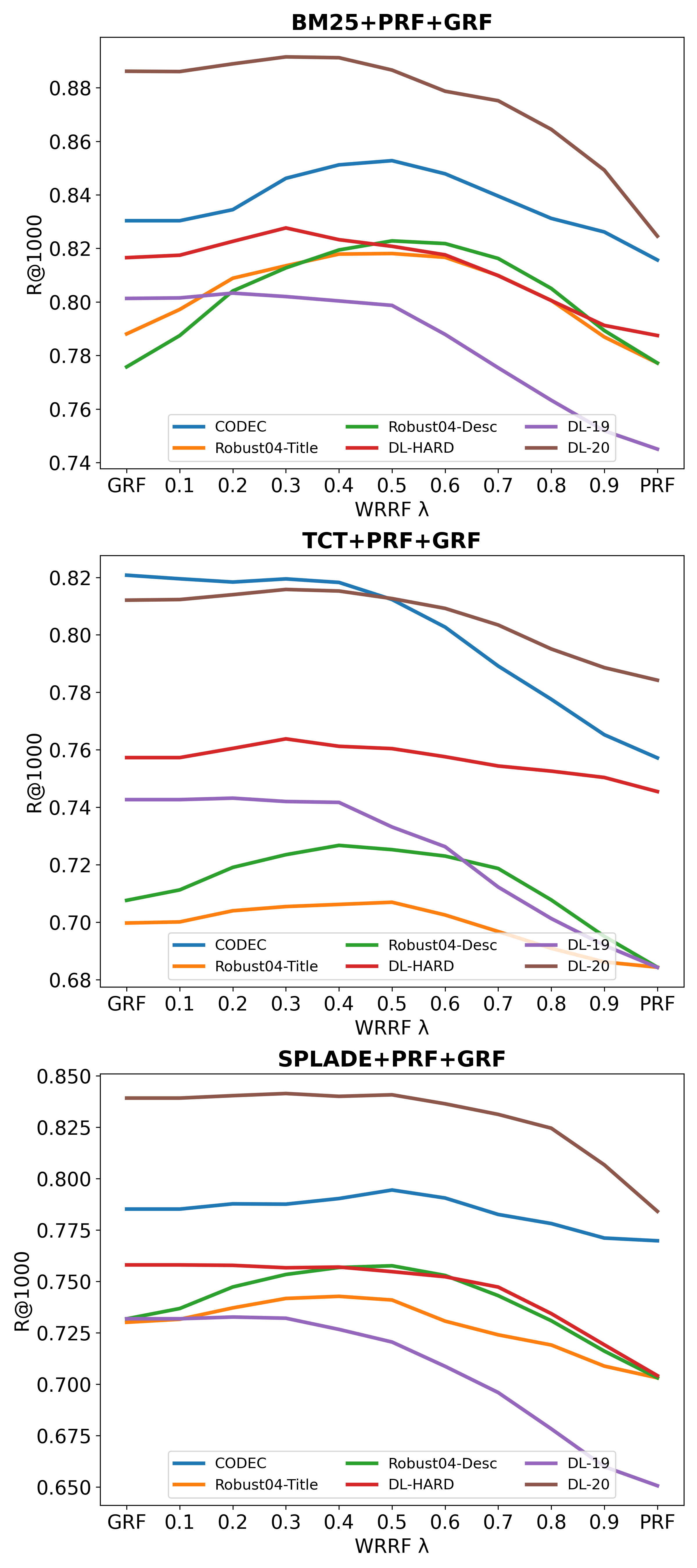}
    \caption{Impact of $\lambda$ on R@1000 of weighted reciprocal rank fusion (WRRF). Where 0.0 is GRF and 1.0 is PRF. 
    }
    \label{img:lambda}
\end{figure}

% --- overview of figire ---
To understand the effect of hyperparameter $\lambda$, Figure~\ref{img:lambda} plots R@1000 of our weighted reciprocal rank fusion method (PRF+GRF) varying $\lambda$, i.e. when $\lambda$ is 0.0 this is the equivalent of GRF, 0.5 equates to RRF~\cite{cormack2009reciprocal}, and 1.0 equates to PRF.
We include all six datasets and our three retrieval paradigms (sparse, dense, and learned sparse).

% --- Findings ---
This graphic highlights that generative and pseudo-relevant feedback methods are complementary. For example, we see R@1000 increases for BM25+PRF+GRF, as $\lambda$ approaches 0.3-0.6, highlighting the benefits of combined ranking signals. Although not as large, we see small, consistent improvements in TCT+PRF+GRF and SPLADE+PRF+GRF. Robust04, in particular, seems to benefit from PRF signals, which may be due to the age of the corpus, i.e. generative documents could be ``too new'' for target relevant documents.
Nonetheless, there are specific datasets, such as DL-19, where PRF incorporation negatively impacts our highly effective GRF model.

% --- Research Question ---
Overall, we show that we can further improve recall by combining the ranking signals of generative and pseudo-relevant feedback models.
We show that GRF and PRF are complementary and explore the impact of weighting generative and pseudo-relevant feedback signals across datasets and models.

\section{Conclusion}
\label{sec:conclusion}

We extend generative-relevance feedback to dense and learned sparse search paradigms.
We find that GRF improves over comparable PRF techniques by around 10\% on both precision and recall-oriented measures.
Furthermore, we conduct query analysis and show that generative and pseudo-relevance feedback have contrasting benefits.
Specifically, generative feedback can provide external context without being dependent on first-pass retrieval, while pseudo-relevance feedback provides information grounded in the corpus.
Based on this, we propose a weighted reciprocal rank fusion method that combines GRF and PRF ranking signals and significantly improves recall over PRF on 95\% of experiments.
We believe this is beginning of a body of work that will combine document retrieval and LLM generation to improve core retrieval effectiveness.   

\section{Acknowledgements}
\label{sec:ack}

This work is supported by the 2019 Bloomberg Data Science Research Grant and the Engineering and Physical Sciences Research Council grant EP/V025708/1.

\bibliographystyle{ACM-Reference-Format}
\balance
%\vfill\eject 
\bibliography{foo}

\end{document}